\documentclass[prd,aps,12pt,nofootinbib,superscriptaddress]{revtex4}
\usepackage{epsfig,graphicx}

\usepackage[dvips]{color}
\usepackage{amsmath}

\begin{document}

\title{Cosmological model with variable vacuum pressure}
%Emergent cosmology: problems and perspectives}

\author{L.~L.~Jenkovszky}\email{jenk@bitp.kiev.ua}
\affiliation{Bogolyubov Institute for Theoretical Physics, Kiev 03680, Ukraine} %

\author{V.~I.~Zhdanov}\email{valeryzhdanov@gmail.com}
\affiliation{Astronomical Observatory, Taras Shevchenko National University of Kyiv} %

\author{E.~J.~Stukalo}
\affiliation{Physical Faculty, Taras Shevchenko National University of Kyiv} %

\begin{abstract}
Scenarios of the cosmological evolution  are studied by using an
equation of state (EoS) having points where the specific enthalpy
of the cosmological fluid vanishes. A large class of barotropic
EoS's admits, depending upon initial conditions, analogues of the
``Big Rip", as well as solutions describing exponential inflation
followed by usual matter dominance; their classification is
proposed. We discuss extensions to more general two-parametric EoS
dealing with a pre-inflationary evolution and yielding stages with
both increasing and decreasing energy density as a function of
time. Possible cosmological scenarios with transitions from
collapse to an expanding Universe or a closed oscillating one,
without reaching a singularity are included.
\end{abstract}


\maketitle

\section{Introduction} \label{Introduction}
The Standard Cold Dark Matter cosmological model with $\Lambda\ne
0$ ($\Lambda$CDM model)  describes a huge amount of observational
data \cite{WMAP,Plank} that determine the cosmological fraction of
the baryonic matter, the cold dark matter and the dark energy. On
the other hand, the $\Lambda$CDM model leads to the well-known
problems of horizon and of spatial flatness (see, {\it e.g.}
\cite{Weinberg,Linde}). These problems are currently resolved by
introducing the inflationary period \cite{inflation} in the early
Universe, leading to a high degree of isotropy of the cosmic
microwave background and spatial flatness. This can be achieved
either within modifications of General Relativity or by taking
into account some additional physical fields \cite{Sahni,Linde},
which are manifest at cosmological scales.

Traditionally, the bulk of matter in the Universe is represented
by a sum of baryonic matter, unknown dark matter (DM) with zero
(or very small) pressure and dark energy (DE). In the
hydrodynamical picture, the total pressure is a sum
$p=p_B+p_{DM}+p_{DE}$. This picture can be complemented by
introducing a cosmological ({\it e.g.}, scalar) field. On the
other hand, at present there are no theoretical arguments that
could single out a unique scalar field potential, a Lagrangian of
a modified gravitation theory etc. In this situation, qualitative
considerations of cosmological scenarios often use a
phenomenological equation of state (EoS) describing on the average
the whole matter  in terms of the local frame energy density $e$
and other thermodynamical parameters (see \cite{eos_models} and
references therein).
 A widely accepted model for the dark
energy is a homogeneous perfect fluid with $p=w e$. Cosmic
acceleration requires that $w<-1/3$; recent WMAP and
Planck results \cite{WMAP,Plank} are consistent with the
value $w \approx -1$, however ``phantom DE",  for which $w<-1$
\cite{BigRip}, is not completely ruled out \cite{phantom}.

A number of generalizations  dealingwith either explicit forms
of EoS or having a parametric form exist \cite{eos_models}. In
this paper we {use an} EoS containing a contribution {from}
ordinary matter with { a} linear part $p=w e$, $w=const$ ({\it
e.g.}, due to ``hot" matter with $w=1/3$) and a nonlinear term
depending on the total energy density $e$. This EoS is inspired by
the well-known bag models of the quark-hadron (de)confinement
phase transition when quarks and gluons coagulate to form hadrons
(see, {\it e.g.}, \cite{Shurik,JKS,JKS1}). These models involve
the so called ``bag" pressure constant $\bar{B}$ that can be
interpreted as the effect of the quantum-chromodynamical vacuum. A
generalization in which the constant $\bar{B}$ is replaced by an
increasing function of temperature $T$, was suggested by
K\"allman \cite{Kallman}. This modification of the EoS was
re-derived and its hydrodynamical and cosmological consequences
have been discussed in a number of papers (see \cite{JKS, JKS1,
ECh} and references therein). The possibility that a small
fraction of colored objects escape hadronization, surviving as
islands of free coloured particles, called quark nuggets was
studied {\it e.g.} in Ref.~\cite{Odessa}.

Possible variation of the cosmological constant with temperature
$T$ or energy density $e$ was discussed {\it e.g.} in Ref.
\cite{Lambda}. Of interest is the connection of this phenomenon
with the variable quark bag ``constant" $B$ considered in the
present paper. Below we extend the idea of the variable vacuum
pressure $B$ to very early times of the Universe, although we do
not exclude that the nonlinear part of the equation of state can
have different origin. We assume that $B$ may be a
 function of $e$. This is applied to equations of the homogeneous isotropic
Universe  (Section \ref{hydrodynamical_model}) dealing with a
barotropic EoS. In Section \ref{section3} we discuss more general
equations of state that involve two thermodynamical parameters,
namely, the energy density and the specific volume. The results
are summarized in Section \ref{conclusions}.

\section{Hydrodynamical models of cosmological evolution}
\label{hydrodynamical_model}
\subsection{Preliminaries: homogeneous isotropic cosmology}
Let us write the Friedmann equations for a homogeneous and
isotropic Universe for the FLRW metric
\begin{equation}\label{metric}
ds^2=dt^2-a^2(t)\left[ d\chi^2+F^2(\chi)dO^2 \right],
\end{equation}
where $a(t)$ is the scale factor, $dO$ is the distance element on
the unit sphere, $F(x)=\sin(x)$ for the closed Universe,
$F(x)=\sinh(x)$ for  open one, and $F(x)=x$ for spatially flat
models; correspondingly, in what follows $k=1,-1,0$ .

The Friedmann equations for the scale factor are
\begin{equation}\label{friedmann_0}
\frac{d^2a}{dt^2}=-\frac{4\pi }{3}a(e+3p),
\end{equation}
where in case of a hydrodynamical cosmological models $e$ stands
for the energy density of all kinds of matter in the Universe and
$p$ is the effective pressure\footnote{We use the system of units in which
$G=1$ and $c=1$.};
\begin{equation}\label{friedmann_1}
H^2=\frac{8\pi}{3}e-\frac{k}{a^2},
\end{equation}
where $H=a^{-1}da/dt$. Here we do not introduce explicitly the
cosmological constant, because it can be incorporated in $e$ and
$p$ as ``dark energy" with EoS $p_{DE}=-e_{DE}$; we remind
that the present-day value of  $\Lambda$ (from $\Lambda$CDM
cosmological model) can be neglected for the redshifts $z\sim 10$
or greater.
 For our purposes it is sufficient
to use only Eq. (\ref{friedmann_1}) as we further use the
hydrodynamical equation
\begin{equation}\label{hydro}
\frac{de}{dt}+3(e+p)H=0,
\end{equation}
or
\begin{equation}\label{hydrolna}
\frac{de}{dX}=-3h=0,
\end{equation}
where $X=\ln a$ and $h=e+p$ is the specific enthalpy. With account
for (\ref{friedmann_1}) we have
\begin{equation}\label{hydro_3}
\frac{de}{dt}=-3S(e+p)\sqrt{\frac{8\pi}{3}e-\frac{k}{a^2}},
\end{equation}
where $S=1$ in the case of the expanding Universe and $S=-1$ in
case of the contracting one. In what follows we assume $S=1$ for
$k=0,-1$.

\subsection{Barotropic equation of state}\label{one-parametric}
The above equations must be complemented by an
EoS. We use the analogy with the quark bag model, where the
bag pressure can be interpreted as the effect of the
quantum-chromodynamical vacuum. Similarly, for energy densities
at very early stages of the cosmological
evolution, we assume that there was a strong negative bag pressure
inherent of the cosmological vacuum for all physical interactions.

In generalizations of the quark bag model \cite{Kallman,JKS}, the
bag constant $\bar{B}$ is replaced by a function of temperature
$T$, so that $p(T)=A_{qg}T^4-B(T)$. In the standard bag model, the
first term of this equation describes the ultra-relativistic gas.
Introduction of temperature as an independent variable is
preferable
 in case of thermodynamical equilibrium, because it enables us to
characterize different components of the cosmological fluid by the
same parameter $T$. On the other hand, in qualitative
considerations of cosmological problems it is  more convenient to
introduce an effective EoS in terms of energy density $e$.

Let us start with the barotropic EoS
 \begin{equation}\label{b-of-e}
 p(e)=we-B(e),
\end{equation}
where we assume that $w>-1$ and the corresponding term represents
an input of an ``ordinary" matter, {\it e.g.}, $w=1/3$  for the
hot matter (ultra-relativistic gas). We write the ``vacuum
pressure" as $B(e)=f(e)e$, and we assume that $f(e)$ is a
monotonically increasing function. We suppose, for simplicity,
that $f(0)=0$, so that for small densities the effect of the
vacuum pressure be negligible \footnote {Consistency with
the $\Lambda$CDM cosmological model for small densities ({\it
e.g.}, for the present era) requires the addition of a positive
constant to $f(e)$.}. Equation (\ref{hydro}) then assumes the form
 \begin{equation}\label{energy-of-a}
\frac{de}{dX}=-3e\,[1+w-f(e)], \quad a>0, \quad X=\ln a.
\end{equation}
The crucial point is the existence of some value $e_0$ where the
specific enthalpy vanishes, {\it i.e.} $e_0: \,\, f(e_0)=1+w$;
this will be asumed in what follows.  We see from (\ref{energy-of-a})
that the behavior of the trajectories in the $a-e$ plane near
$e_0$ do not depend on the choice of $k$. However, contrary to the
cases of $k=0$ and $k=-1$, for $k=1$, the trajectories cannot be
always extended to all positive values of $a$, and this requires
additional considerations (see below).

 By using (\ref{hydro}) and (\ref{friedmann_1}) we get the equation
 \begin{equation}\label{energy-of-t}
\frac{de}{dt}=-3Se\,[1+w-f(e)]\sqrt{\frac{8\pi
}{3}e-\frac{k}{a^2}}.
\end{equation}

\begin{figure}
\centerline{\includegraphics[width=4.0in]{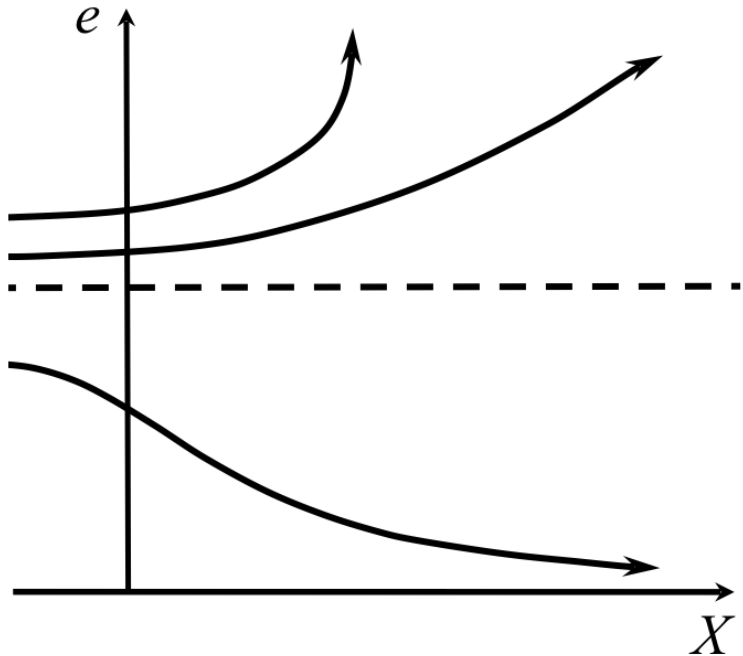}} \caption{{\small
Qualitative behavior of solutions of equation (\ref{energy-of-a})
for $k=0$ and $k=-1$; $X=\ln a$, $S=+1$, monotonically increasing
$f(e)$. The trivial solution $e=e_0$ is shown by a dashed line.
The types of qualitative behavior from bottom to top are: {\bf A1,
A2, A3}.}} \label{Fig_1}
\end{figure}

We now describe types of the qualitative behavior of solutions
 $e(t),a(t)$ of equation (\ref{energy-of-a}) in more details. For $k=0$ they are illustrated in Fig. \ref{Fig_1} in case of the expansion (S=1)
 of the spatially flat Universe.

{\bf A1}:  The solutions are represented by the lower curve in
Fig. \ref{Fig_1} lying completely in the domain of variables
$(e,X)$ such that $e \in (0, e_0)$. For $X\to-\infty\quad (a\to
0)$ we have $e\to e_0$, $H\approx const>0$ and this corresponds to
a long period of an exponential inflation.  As $t$ increases, the
energy density decreases, and the contribution of the vacuum
pressure becomes negligible. This is, however, the consequence of
condition $f(0)=0$ and it can be easily corrected if we want to
take into account the present value of $\Lambda\ne 0$.

{\bf A2}: The solutions are represented by the curve lying in the
second region, $e>e_0$, between the other curves in Fig.
\ref{Fig_1}. For such a solution $e(X)$ is defined for all $X$,
this function is monotonically increasing up to infinity;
 $a(t)$ and $e(t)$ are defined for all $t>0$.
This scenario takes place, {\it e.g.}, in case of a bounded
$f(e)$.

{\bf A3}: The solutions are represented by the top curve in Fig.
\ref{Fig_1} lying in the region $e>e_0$, with a ``runaway"
behavior like that of the Big Rip \cite{BigRip}. A solution of
this type blows up at some finite time, and it cannot be extended
for all $X$ and/or all $t$. Such a behavior\footnote{To save the
space we show the trajectories corresponding to different kinds of
$f(e)$ in the same figure.} occurs, {\it e.g.}, if we suppose that
$f(y)$ grows faster than $\sim y^{\epsilon}$, ${\epsilon}>0$ for
large $y$.

An example with explicit EoS and analytic solutions is given in
Appendix \ref{Analytic solutions}.

For $k=-1$ we always have $8\pi e/3 -k/a^2 >0$; therefore the
qualitative situation is  analogous to the previous case. For
$X=\ln a \to -\infty$ ($a\to 0$) we have the same dependence
$e(X)$ described by Fig. \ref{Fig_1} because Eqs. (\ref{hydrolna})
and (\ref{energy-of-a})  do not contain $k$. There can be either
{\bf A1, A2} or {\bf A3} as possible cases for $e(X)$, though they
can yield different asymptotical behavior for $a(t)$ and $e(t)$
for $t\to 0$ or $t\to\infty$.

It is easy to see, in virtue of (\ref{friedmann_1}), that, since
$e\to e_0$ as $t\to 0$, we have $a(t)\approx t$. In the region $e\in
(0, e_0)$ all the solutions $e(t)\to 0$ are monotonically
decreasing functions and $a(t)\to \infty$ for $t\to \infty$. In
case of a fine tuning, if, at some $t_0$, we have $e(t_0)$
sufficiently close to $e_0$, this closeness will remain during a
long time; then in virtue of (\ref{friedmann_1}) $\dot X>q$, where
$q\sim\sqrt{8\pi e_0/3}$, whence $a(t)>a(t_0)\exp[q(t-t_0]$.
 In the second region, $e>e_0$, the situation is also the
same as for $k=0$: we  have either the ``runaway" solutions {\bf
A3} or solutions {\bf A2}.

 In both cases, $k=0$ and $k=-1$, the line $e=e_0$ is a
solution; other solutions (either from the region $e<e_0$ or from
the region $e>e_0$), cannot cross this line\footnote{It would
contradict the uniqueness of the solution of (\ref{energy-of-a})
with initial condition $e=e_0$ at some initial point.}.

\begin{figure}
\centerline{\includegraphics[width=4.0in]{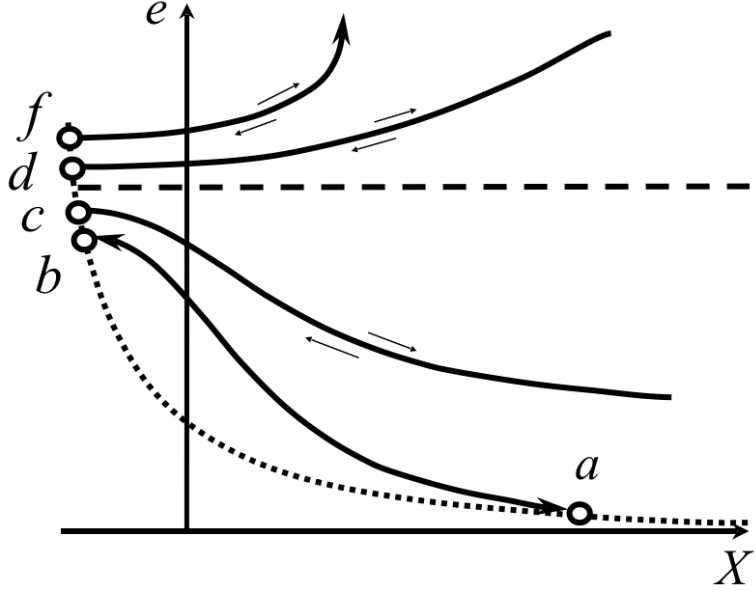}} \caption{{\small
Qualitative behavior of solutions of (\ref{energy-of-a}) for
$k=1$, $w\in (-1,-1/3)$, with monotonically increasing $f(e)$. The
dashed line corresponds to solution $e=e_0$, the dotted curve is
defined by equation  $\Xi(e,X)=0$. The possible types of
trajectories are (from the bottom to the top): {\bf A4}
(oscillating between $a,b$), and {\bf A5, A6, A7} with turning
points $c,d,f$ of $X(t),$ correspondingly. The arrows near the
trajectories starting at $c,d,f$ indicate two possible directions
of motion along these trajectories; here contraction ($S=-1$) is
followed by expansion ($S=+1$).}} \label{Fig_2}
\end{figure}

$k=1$: Since the right hand side of equation
(\ref{friedmann_1}) must be positive, the trajectories are located
to the right of the curve $\Xi(e,X)\equiv 8\pi e/3-\exp(-2X)=0$.
The intersections of  solutions $e(X)$ of equation
(\ref{energy-of-a}) with this curve typically correspond to simple
zeros\footnote{This is a simple zero unless at this point
$e+3p=0$, i.e. $d^2a/dt^2=0$ simultaneously with $da/dt=0$. In the
latter case the solution spends an infinite time near this point.}
of function $\Xi(e(X),X)$; therefore, they describe turning points
of solutions $a(t)$ to equation (\ref{friedmann_1}) that can be
rewritten as
\begin{equation}\label{Xi}
\left(\frac{dX}{dt}\right)^2=\Xi(e(X),X), \quad X=\ln a.
\end{equation}
Any turning point in its neighborhood  relates two branches of the
cosmological evolution with different signs of $S$: from a
contracting ($S=-1$) to an expanding $(S=1$) Universe. The mutual
arrangement of the trajectories of  (\ref{energy-of-a}) in the
$e-X$ plane (Fig. \ref{Fig_2}) shows that there must be at least
one intersection of any trajectory with the curve of turning
points $\Xi(e,X)=0$. Correspondingly, possible types of solutions
are as follows.

{\bf A4}: Oscillating solutions in the domain $e<e_0$ for
$w>-1/3$. On account for $f(e)\to 0$ as $a\to \infty$, due to
equation (\ref{energy-of-a}), we have $e\sim a^{-3(1+w)}$.
Therefore, for a solution that starts, {\it e.g.} at the turning
point $b$ of the solution $X(t)$, there is necessarily another
turning point ($a$) (see Fig. \ref{Fig_2}). There is then an
oscillating solution of ($S=\pm 1$) described by the curve between
points $a$ and $b$; in this model $e(t)$ oscillates without
reaching any singularity.

For $w\in (-1,-1/3)$,  depending on initial conditions, one can
also have  an oscillatory behavior like {\bf A4}.  On the other
hand, in this case an alternative version with ever expanding
Universe is possible:

{\bf A5}: monotonically decreasing solutions $e(X)$, such as those
starting from the turning point $c$ in Fig. \ref{Fig_2} and
existing for all $X$ to the right of the turning point $c$. Here a
contraction ($S=-1$) is followed by expansion ($S=1,\,\,X(t)\to
\infty$ as $t\to \infty$); the singularity is not reached and $e$
remains finite for any time.

In the region $e>e_0$ we have a monotonically increasing solution
$e(X)$ of equation (\ref{energy-of-a}). Similarly to {\bf A2} and
{\bf A3}, we have two possibilities.

{\bf A6}: Solutions that are represented by trajectories that have
a turning point (like $d$ in Fig.\ref{Fig_2}) and tend to infinity
as $X\to \infty$. For $e(t)$ and $a(t)$, as functions of $t$, we
have transition from contraction to expansion at $d$.

{\bf A7}:  Solutions with a transition from contraction to
expansion at some turning point ($f$ in Fig.\ref{Fig_2}, upper
curve); these solutions blows up at some finite $X$.

Ending  this Section, we note that if we have two zero points of
enthalpy\footnote{provided we do not assume $f(e)$ to be
monotonic.}, i.e.  $e_0$ and $e_1$ such that $f(e_0)=1+w$ and
$f(e_1)=1+w$; consequently we get a family of solutions with $e\to
e_0$ for $a\to 0$ and $e\to e_1$ for $a\to \infty$. In this case
$e_1$ can be interpreted as a modern value of the dark energy
density described by $\Lambda\ne 0$.

\subsection{Monotonically decreasing $f(e)$}\label{one-parametric_rev}
\begin{figure}
\centerline{\includegraphics[width=4.0in]{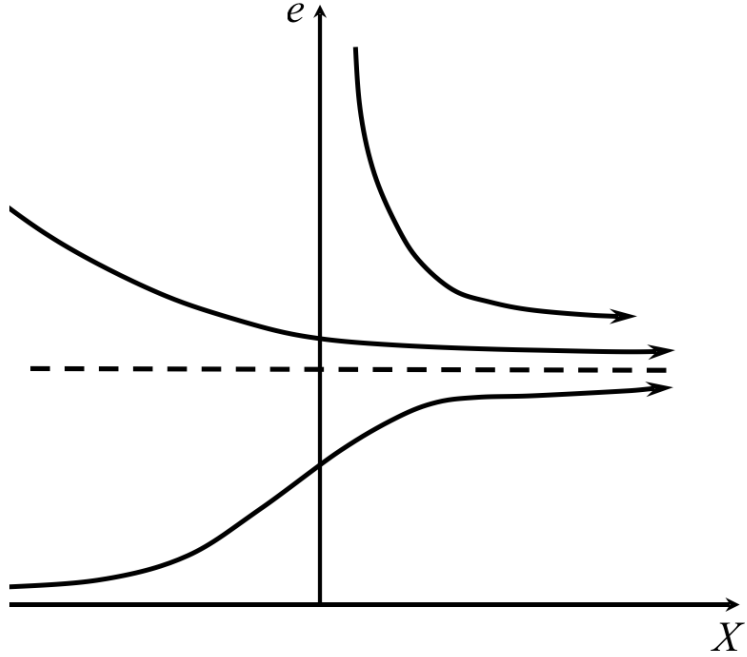}} \caption{{\small
Types of solutions of (\ref{energy-of-a}) for $k=0$ and $k=-1$;
$X=\ln a$, $S=+1$, monotonically decreasing $f(e)$. The dashed
line is $e=e_0$. From bottom to top: {\bf B1, B2, B3}.}}
\label{Fig_3}
\end{figure}

\begin{figure}
\centerline{\includegraphics[width=5.0in]{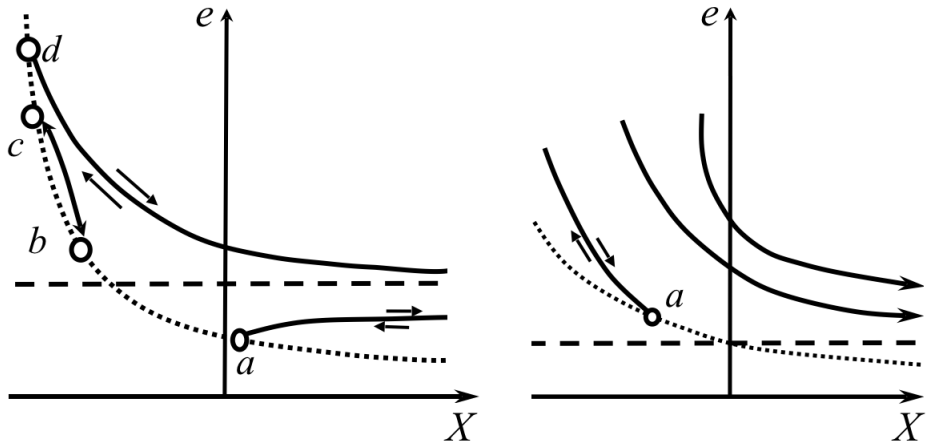}} \caption{{\small
Types of solutions of (\ref{energy-of-a}) for $k=1$; $X=\ln a$,
monotonically decreasing $f(e)$. The dashed line corresponds to
$e=e_0$, the dotted curve is defined by relation $\Xi(e,X)=0$. The
left panel represents trajectories ({\bf B4--B6}) with bounded
energy density, the right panel -- with unbounded one ({\bf
B7--B9)}.}} \label{Fig_4}
\end{figure}

For completeness, we consider also the case of monotonically
decreasing function $f(e)$ with the same relation for
$h=e[1+w-f(e)]$; we suppose that there exists $e_0$ such that
$f(e_0)=1+w$ and therefore $h(e_0)=0$. Then in the domain
$\{(e,X): e>e_0\}$ we have $de/dX<0$ along the trajectories, and
for $e\in(0,e_0)$ we have $de/dX>0$. The behavior of the
trajectories is shown in Figs. \ref{Fig_3}, \ref{Fig_4}. The
qualitative types of the solutions of (\ref{energy-of-a}) are as
follows.

{\bf B1} ($k=0,-1$):  Solutions in the domain $\{(e,X): e<e_0\}$:
$e(X)$ defined on $(-\infty, \infty)$, $e(X)\to 0$ for $X\to
-\infty$, and  $e(X)\to e_0$ for $X\to \infty$.

{\bf B2} ($k=0,-1$):  Solutions defined on $(-\infty, \infty)$ and
we have $e(X)\to \infty$ for $X\to -\infty$, and $e(X)\to e_0$ for
$X\to \infty$. This is possible, {\it e.g.}, in case of a bounded
$f(e)$; for $3[1+w-f(\infty)]>2$ we have the same asymptotic
behavior of $a(t)$, $t\to 0$ both for $k=0$ and $k=\pm 1$. This
kind of behavior is appropriate for the classical $\Lambda$CDM
cosmological models with $p=0$ or $p=e/3$ (see, {\it e.g.}
\cite{Weinberg,Dinverno}).

{\bf B3}: Solutions $e(X)$  defined for $\infty<X_1 <X< \infty$,
where $e(X)\to \infty$ as $X\to X_1+0$ ("Big Rip in the past",
upper curve in Fig. \ref{Fig_3}).  For $k=0,-1$ this type is
possible, {\it e.g.}, if $f(y)<0$ for $y\to \infty$ and $|f(y)|$
grows faster than $\sim y^{\epsilon}$, ${\epsilon}>0$. For $X\to
\infty$ we have $e(X)\to e_0$.

The next six types of solutions of (\ref{energy-of-a}), shown in
Fig. \ref{Fig_4}, deal with the closed Universe ($k=1$); {\bf B4,
B5, B6}  describe solutions with bounded energy density (Fig.
\ref{Fig_4}, left panel)  and {\bf B7, B8, B9} are solutions with
unbounded $e$ (Fig. \ref{Fig_4}, right panel).

{\bf B4} ($k=1$): Monotonically increasing $e(X)$ in the domain
$e<e_0$ with turning point ($a$).

{\bf B5} ($k=1$): Oscillating  solution (between $b$ and $c$).

{\bf B6} ($k=1$): Monotonically decreasing $e(X)$ in the domain
$e>e_0$ with turning point ($d$).

Fig. \ref{Fig_4}, right panel:

{\bf B7} ($k=1$): Unbounded monotonically decreasing $e(X)$ in the
domain $e>e_0$ with turning point $a$; $e(X)\to \infty$ as $X\to
-\infty$. The Universe starts with infinite $e$, whereupon
expansion is replaced by contraction after passing the turning
point $a$.

{\bf B8} ($k=1$): Unbounded monotonically decreasing $e(X)$ in the
domain $e>e_0$, defined for all $X$;  $e(X)\to \infty$ as $X\to
-\infty$; $e(X)\to e_0$ as $X\to \infty$.

{\bf B9} ($k=1$): Unbounded monotonically decreasing $e(X)$ in the
domain $e>e_0$, for $X>X_1>-\infty$;  $e(X)\to \infty$ as $X\to
X_1+0$; $e(X)\to e_0$ as $X\to \infty$.

Note that one can easily extend  the results of subsections
\ref{one-parametric} and \ref{one-parametric_rev} to the case of
negative $f(e)$ and $1+w<0$. In this case we have the same types
of trajectories as those shown in Figs. \ref{Fig_1}--\ref{Fig_4}.

\section{Generalization: two-parametric EoS}\label{section3}
 In the preceding section we have used the fact that the solution $e(x)$ of
the first order differential equation (\ref{energy-of-a}) cannot
 cross the line $e\equiv e_0$, since this is also a
solution of (\ref{energy-of-a}). However, it is interesting to
consider scenarios when the sign of $de/da=0$ may change during
the cosmological evolution. To this end we relax the conditions of
the preceding section and assume that the EoS depends on two
independent thermodynamical variables. On the other hand, one can
expect that, if the dependence of the pressure on the additional
variable, besides $e$, is weak, the cosmological evolution will be
very close to that of the preceding Section and a solution that
crosses zero points of $h$ will spend a considerable time near
these points (and thus can invoke a kind of the exponential
inflation).

In this Section, instead of Eq. (\ref{energy-of-a}), we
prefer to deal directly with Eq. (\ref{hydrolna}), where
$p=p(e,v),\,\,h=h(e,v)$, $v=1/n$, $n$ is the proper frame baryon
number density.  Due to baryon conservation, in case of metric
(\ref{metric}), we have
\begin{equation}
\label{baryon_cons} v=v_0(a/a_0)^3,
\end{equation}
where $v_0,\,a_0$ are the values at some $t_0$.

Instead of the straight line $e=e_0$,  we suppose that there
is (only) one curve $e=E(X)>0$ where the specific enthalpy
$h(e,X)$ vanishes.
 Obviously, in the general case, the function
$e=E(X)$ does not satisfy (\ref{hydrolna})  and some trajectories
of solutions can cross this curve.  The most simple extension of
the corresponding assumptions of  Section \ref{one-parametric} is
that  the function $E(X)$ is defined for all $X$ and it is a
monotonic function. The trajectories slightly differ for different
signs
 of monotonicity. There are four possibilities: ({\bf C1}) $E(X)$ is
 monotonically increasing, $h(e,X)<0$ for $e>E(X)$; ({\bf C2}) the same
 sign of monotonicity, but $h(e,X)>0$ for $e>E(X)$; ({\bf C3}) $E(X)$ is
 monotonically decreasing, $h(e,X)<0$ for $e>E(X)$; ({\bf C4}) the same
 sign of monotonicity as in ({\bf C3}), but $h(e,X)>0$ for $e>E(X)$.

\begin{figure}
\centerline{\includegraphics[width=5.0in]{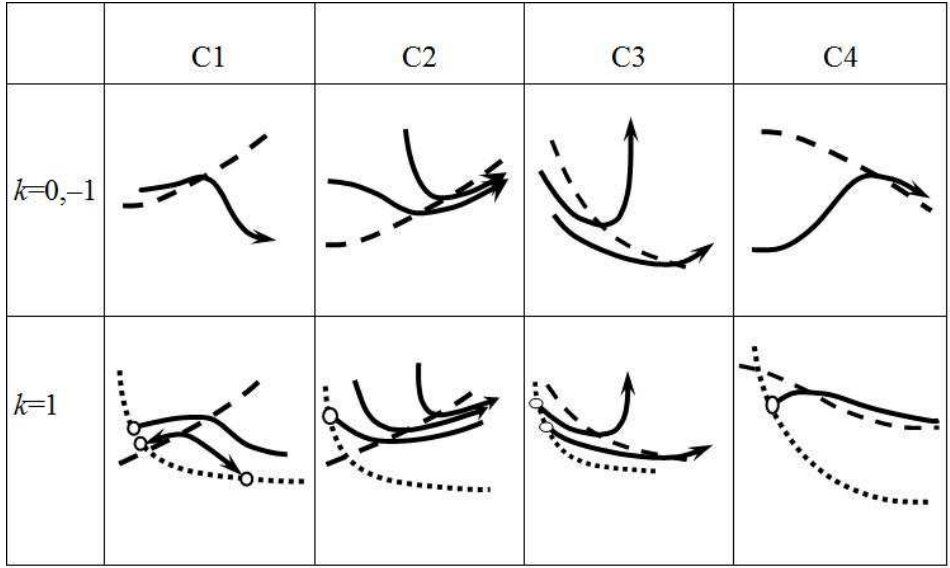}}
\caption{Solutions of equation (\ref{hydrolna}), that change the
sign of $de/dX$: possible qualitative behavior in $e-X$ plane. On
dashed curve, $de/dX=0$. On the upper panel trajectories in case of
$k=0,-1$, $S=+1$ are shown. On the lower panel ($k=1$, closed
Universe) small circles on the dotted curve  indicate
 turning points $\Xi(e,X)=0$ similar to those of the previous Section.
} \label{Table}
\end{figure}

Possible trajectories of the system
(\ref{friedmann_1})-(\ref{hydrolna}) that do not cross the curve
$e=E(X)$  are qualitatively  the same as those of Section
\ref{hydrodynamical_model}, so we concentrate on types of
trajectories for which the sign of $de/dX$ changes, as shown
schematically in Fig. \ref{Table}.  These scenarios of
cosmological evolution differ considerably from the standard
pictures described in the textbooks (see, {\it e.g.},
\cite{Dinverno}, Fig. 23.1). The trajectories that do not cross
the curve $h=0$ are not shown in Fig. \ref{Table}; for example, in
case of {\bf C1}, $k=0,-1$ there may be infinite solutions  that
move to the area above the curve $h=0$.

In any case, for $k=0,-1$ the initially expanding Universe will
expand forever. In case of the closed Universe ($k=1$), for some
solutions, there is  a possibility of return to a contraction;
there can be also periodic solutions and solutions with return
from contraction to expansion.

It is interesting to note that there are scenarios when $e(t)$
increases and then decreases either to zero or to some constant
value (see Fig. \ref{Table}, {\bf C1} and {\bf C4}). Any more
complete description may be obtained only if more information
about $h(e,X)$ is provided. Two concrete examples with types {\bf
C1} and {\bf C4} are presented in Appendix \ref{two_examples}.

\section{Conclusions}\label{conclusions}
We have investigated a  class of nonlinear equations of state that
have points of zero enthalpy; to do so, we used
 general  qualitative properties known from the theory of
ordinary differential equations. In the case of the barotropic EoS
occurrence of such point leads
 to the possibility of a period of rapid expansion with almost constant energy
 density. Under appropriate initial conditions, this period can last as long
as needed to provide the necessary size of the causally connected
 regions in future, ensuring, {\it e.g.}, an adequate solution to the
well-known problem of horizon etc.

Under general assumptions, we have classified possible
cosmological scenarios, including a  number of those without
 the cosmological singularity. In case of an open and spatially
flat metrics, the Universe expands either forever or undergoes a
Big Rip. In case of the closed Universe there can be also
transitions from collapse to expansion or vice versa. A closed
oscillating Universe without reaching a singularity is also
possible.

Our findings deal with  simple EoS allowing to
discuss different scenarios of the early cosmological evolution.
Nevertheless, if we assume that these considerations have
relevance to reality, one can speculate on why our Universe starts
with small deviations from the state with zero enthalpy, for
example due to quantum fluctuations near this state. These
fluctuations can start either ``phantom" cosmological evolution
leading to a kind of the Big Rip, or ``normal" expansion  with a
transition to the standard $\Lambda$CDM  model described in
textbooks (see, {\it e.g.}, \cite{Weinberg}). To have a
sufficiently long period of
 inflation, a kind of fine tuning is required ({\it e.g.},
approach to the point $h=0$), compelling  to invoke a sort of the
anthropic principle. However, such a tuning is not more
restrictive than, {\it e.g.}, the use of  equation of state $p=w
e$ with $w=-1$.

The above-mentioned requirement of the fine tuning is relaxed if
we consider a two-parametric EoS. In this case, an infinity of
solutions crossing states with zero enthalpy may exist; and there
is an immense freedom in choosing the evolution as the scale
factor $a\to 0$. In case of a sufficiently weak dependence of
$p(e,v)$ on $v$ it is natural to assume that the behavior of
trajectories near points of zero enthalpy will be similar to that
of Section \ref{one-parametric} and there is a sufficiently long
period of inflation.

Furthermore, one may speculate about a multi-component fluid. We
know that at present some form of DE dominates, but moving back in
time its fraction becomes negligible (within the framework of the
$\Lambda$CDM model). On the other hand, for very early
(post-Planckian) times, the other form of  DE must have dominated to
provide inflation. Within such a
 scenario, one may have different inflation epochs
corresponding to the domination of different DE components (cf.
multiple stages of inflation in Ref. \cite{schism}). This is
consonant with the idea of the existence of a ``mini-inflation''
due to the existence of metastable states in strongly interacting
matter \cite{JKS,JKS1}, see also \cite{Shurik}. However, any
smooth interpolation between these different inflation epochs
remains a challenge for the theory.

\section*{Acknowledgments}
We thank the referee for her/his helpful
comments. We are grateful to G.~Stelmakh for his help in preparing our figures.
L.~J. was supported by the National Academy of Sciences
of Ukraine, Department of Astronomy and Physics programme ``Matter
Under Extreme Conditions". V.I.Z. was supported by the Taras
Shevchenko National University of Kyiv, scientific program
``Astronomy and Space Physics".

\vskip 5 cm

\vfill

\appendix

\section{``Equivalent" scalar field
potential}\label{Analytic solutions} It is important to have a
scalar field analogue of the hydrodynamical models in question.
Here we present an analytic example dealing with a special form of
the EoS and  an   "equivalent" scalar field description yielding
the same evolution of the scaling factor.

It is well known that in case of the homogeneous isotropic
cosmology, one can find a scalar field Lagrangian that mimics the
hydrodynamical evolution (see, {\it e.g.}, \cite{Zhdanov,
Odintsov2, Chavanis} and references therein). Here we consider the
scalar field equations giving the same evolution of the
``observable" space-time geometry, {\it i.e.}
 the same dependence $a(t)$, as that following from EoS
$p=p(e,v)$. Consider the scalar field Lagrangian with the standard
kinetic term
\begin{equation}
\label{lagrangian}
L=\frac{1}{2}\,g^{\mu\nu}\varphi_{\mu}\,\varphi_{\nu} -V(\varphi).
\end{equation}
The corresponding energy-momentum tensor is
\begin{equation}
\label{tensor} T_{\mu\nu}=\varphi_{\mu}\,\varphi_{\nu}-g_{\mu\nu}
V(\varphi).
\end{equation}
In case of a homogeneous isotropic cosmology with the metric
(\ref{metric}), it has the form of the hydrodynamical
energy-momentum tensor with the following energy density and
pressure:
\begin{equation}\label{Vphi}
e=\frac{1}{2}\dot \varphi^2+V,\quad p=\frac{1}{2}\dot \varphi^2-V,
\quad \to \quad \dot \varphi^2=e+p,\quad V=\frac{1}{2}(e-p).
\end{equation}
Combining Eqs. (\ref{Vphi}) with Eq.
(\ref{friedmann_0},\ref{friedmann_1}) we obtain a parametric
representation of  the ``equivalent" scalar field potential:
\begin{equation}\label{mimicry0}
V=\frac{1}{4\pi}\left[H^2+\frac{1}{2a}\frac{d^2 a}{dt^2}
+\frac{k}{a^2} \right],
\end{equation}
\begin{equation}\label{mimicry1}
\varphi=\pm \int {\left\{\frac{1}{4\pi}\left[
H^2-\frac{1}{a}\frac{d^2 a}{dt^2} +\frac{k}{a^2}
\right]\right\}^{1/2}dt}.
\end{equation}

 In the case of barotropic EoS, such as (\ref{b-of-e}),
  and spatially flat Universe ($k=0$), it is more convenient to look for a parametric
  representation  of the form   $V=V(e)$, $\varphi=\varphi(e)$.
  Combining $d\varphi/dt$ with
  Eq.(\ref{hydro}) we get $d\varphi/de$ whence
\begin{equation}\label{mimicry}
\varphi=\pm \int \frac{de}{\sqrt{24\pi e(e+p)}} \, ,
\end{equation}
 yielding the potential
$V(\varphi)$ in a parametric form.

There is a number of analytic examples describing the cosmological
evolution in various dynamical DE models (see, {\it e.g.}
\cite{eos_models, Odintsov2, Chavanis}). Below we use particular
examples to illustrate the general statements of the previous
sections.

In case of $f(\xi)\sim \xi^{\,\mu}$, $\mu>0$ we have an EoS of the
form
\begin{equation}\label{EoSanalytic}
p(e)=e\,[w - (1+w)(e/e_0)^{\,\mu}].
\end{equation}
In this case equation (\ref{energy-of-a}) with the initial
condition $e=e_1$ for $a=a_1$ can be easily integrated:
\begin{equation}\label{anal_e_of_a}
e=e_0\,\left[1+A\left(\frac{a}{a_1}\right)^{3\mu(1+w)}\right]^{-1/{\mu}},
\quad A=\left(\frac{e_0}{e_1}\right)^{\mu}-1.
\end{equation}
It is also possible to find an analytic form of $V(\varphi)$.
Equation
 (\ref{mimicry}) yields
\begin{equation}\label{anal_phi}
\varphi=\pm\frac{1}{[6\pi(1+w)]^{1/2}{\,\mu}}\ln{\left(\xi+\sqrt{\xi^2-1}\right)}
+ \varphi_0, \quad \xi=\left(e_0/e\right)^{{\,\mu}/2}
\end{equation}
where $\varphi_0$ is an integration constant. The inverse function
$e=e(\varphi)$ is
\[
e=e_0[\cosh(\pm \,\Phi)]^{-2/{\,\mu}}, \quad \Phi=\mu\,\sqrt{
6\pi(1+w)}\,(\varphi-\varphi_0) .
\]
Substitution into (\ref{Vphi}) yields
\begin{equation}\label{anal_V}
V(\phi)=\frac{e_0}{\left( \cosh \Phi \right) ^{2(1+1/\mu)}}
        \left( 1 +  \frac{1-w}{2}  \sinh^2 \Phi \right).
\end{equation}

This is the potential that ensures the same dependence of the
scale factor $a(t)$ as the EoS (\ref{EoSanalytic}).

\section{Two examples}\label{two_examples}
\begin{figure}
\centerline{\includegraphics[width=4.0in]{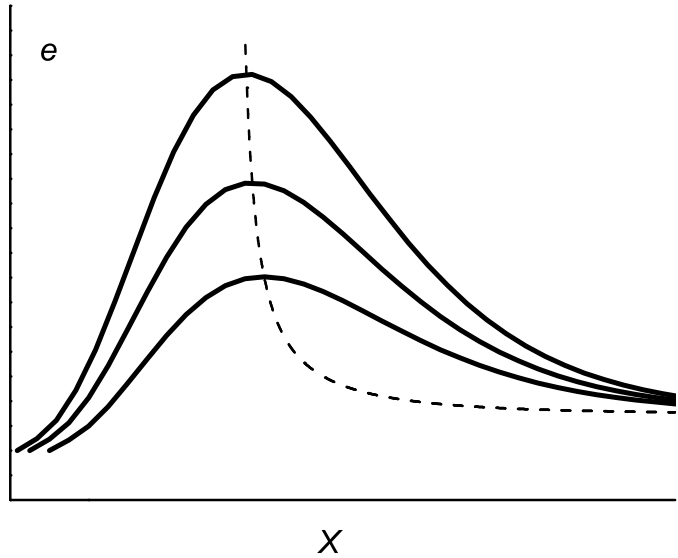}}
\caption{{\small Solutions of (\ref{hydrolna}) corresponding to
(i) crossing the line $de/dX=0$. The choice of the constants is:
$w=1/3$, $f_0=1, \,\,\beta=0.01$. The dashed line is
$E(X)=b_0/[1+w-\beta \exp(-3X)]$}.} \label{fig7N}
\end{figure}

\begin{figure}
%\centerline{\includegraphics[scale=1.0]{Fig.eps}}
\centerline{\includegraphics[width=4.0in]{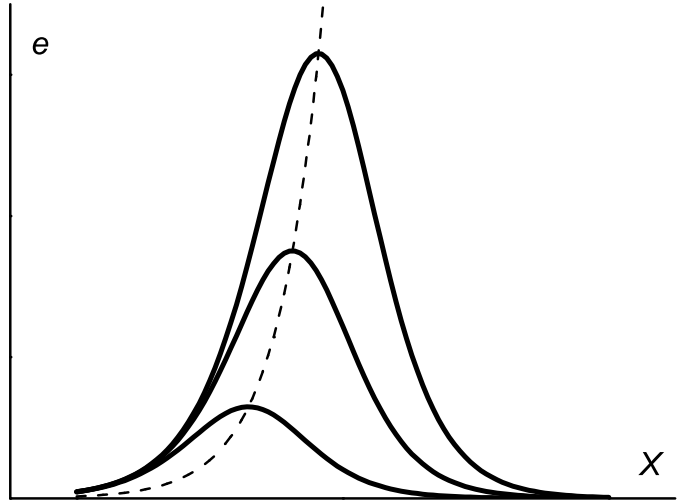}}
\caption{{\small Solutions of (\ref{hydrolna}) corresponding to
(ii), crossing the line $de/dX=0$. The dashed line
 is $E(X)=(1+w)^2/\beta^2 \cdot\exp(6X)$, $\beta=1,w=1/3$}.} \label{fig6N}
\end{figure}

Lacking reliable knowledge about the equation of state in the very
early epoch, below we consider, for illustration purposes, simple
examples of {\bf C1} and {\bf C4} types, $k=0,-1$. The solutions
for $k=1$ can be easily analyzed  using these pictures by
superimposing the line $\Xi(e,X)=0$ onto the graph and  taking
into account the turning points. Assuming Eq. (\ref{b-of-e}), we
set $B(e,v)=b_0(e)+b_1(e)/v$. Then $h(e,X)=(1+w)e-b_0(e) -b_1(e)
\exp(-3X)$. The solutions $e(X)$ of equation (\ref{hydrolna}) are
calculated for most simple cases: (i) $f_0=const>0$, $b_1=\beta
e,\,\, \beta= const>0$ (Fig. \ref{fig7N}); (ii) $b_0=0$,
$b_1(e)=\beta e^{3/2}$ (Fig. \ref{fig6N}). On the figures we show
only those trajectories that cross the line $h(e,X)=0$ where
$de/dX$ changes its sign.

In case of (i)   all solutions $e(X)$ crossing the curve $E(X)$,
tend to $E(\infty)=b_0/(1+w)$ for $X\to \infty$. Besides, there
are monotonically increasing solutions (not shown in Fig.
\ref{fig7N}) below the curve with the same asymptotic behavior for
$X\to \infty$. For large negative $X$, the solutions lead to
nonphysical values $e<0$.

In case of (ii)  all solutions $e(X)$ after crossing the curve
$E(X)$ tend  to zero for $X\to \pm\infty $. However, there are
also solutions (not shown in Fig. \ref{fig6N}) to the left of the
curve tending to infinity for $X\to X_1-0$ at some finite $X_1$.

\end{document}